\documentclass[12pt]{iopart}

\usepackage{iopams}  
\usepackage{graphicx}
\begin{document}

\title{Thermo-optical response of photonic crystal cavities operating in the visible}

\author{Janik Wolters$^{1*}$, Niko Nikolay$^{1}$, Max Schoengen$^{2}$, Andreas W. Schell$^{1}$, J\"urgen Probst$^{2}$, Bernd L\"ochel$^{2}$, and Oliver Benson$^{1}$} 
\address{$^{1}$Nano-Optics, Institute of Physics, Humboldt-Universit\"{a}t zu
Berlin, Newtonstr.~15, D-12489  Berlin, Germany\\$^{2}$Department for Micro- and Nanostructured Optical Systems, Helmholtz-Zentrum Berlin f\"{u}r Materialien und Energie GmbH,
Albert-Einstein-Str. 15, D-12489 Berlin, Germany}

\ead{* janik.wolters@physik.hu-berlin.de}

\begin{abstract}
In this paper we study thermo-optical effects in gallium phosphite photonic crystal cavities in the visible. By measuring the shift of narrow resonances we derive the temperature dependency of the
local refractive index of gallium phosphide in attoliter volumina over a temperature range between 5 K and 300 K at a wavelength of about 605 nm. 
Additionally, the potential of photonic crystal cavities for thermo-optical switching of visible light is investigated. As an example we demonstrate thermo-optical switching with 13 dB contrast.

\end{abstract}

\maketitle

\tableofcontents
\title[Sensitive Local Response of High-Q Photonic Crystal Cavities]{}

\section{Introduction}

Photonic crystal (PC) nano cavities have increasing importance for quantum optics, photonics, and sensing applications~\cite{Vahala2003}. 
The main reason for this is their ability to confine the electromagnetic field in volumina comparable to the cubic wavelength and thus to strongly enhance light-matter interaction~\cite{Akahane2003, OBrien2009}. 
Using PC cavities not only Purcell enhancement of single quantum emitters could be shown~\cite{Wolters2010,Englund2010, Englund2007b}, but also the strong coupling regime of cavity quantum electrodynamics was reached with single quantum dots in PC cavities~\cite{Englund2007,Faraon2011,Yoshie2004}. 
With such systems, lasing oscillations of a single quantum dot~\cite{Nomura2010} and ultrafast all-optical switching by sub-femtojoule pulses~\cite{Taniyama2010} and single photons could be demonstrated recently~\cite{Volz2011}.

The high quality factor of PC cavities corresponds to a narrow resonance line. Its shift is a very sensitive indicator for local modifications of the optical properties. A prominent example of such modifications is the dependency of the index of refraction on the local temperature. Cooling of PC cavities is mandatory for many quantum optics experiments, since emission rate and coherence of most quantum emitters improve drastically \cite{Henneberger2009}. However, for complex integrated circuits \cite{OBrien2009,Englund2007,Benson2011} operating at cryogenic temperatures as intended, the shifts of the resonance wavelength and modifications of quality factors \cite{Joannopoulos2007} of individual elements have to be compensated. Therefore, design and material properties should be known a priori. In this paper we first investigate in detail the resonance shift of a PC cavity from room temperature to 5 K. From this we derive the index of refraction of the cavity material (GaP) as a basis for a predictable design of PC structures in the visible spectral range.   

In a second part of the paper local heating and its effect on the PC cavity resonances is investigated. We perform experimental studies of thermo-optical switching of visible light. Based on theoretical analysis we estimate that due to the ultra-small volume of PC cavities even thermal effects may occur with very short time constants. 

\section{Photonic crystals in gallium phosphide}
The photonic crystal structures used throughout our experiments are made of GaP and operate in the visible from 600~nm to 650~nm ~\cite{Rivoire2008,Wolters2010}.
In this material thermal expansion can be neglected and the only relevant process when changing the temperature is the change of the refractive index $n$~\cite{Pikhtin1977,Slack1975}. 
This effect is exploited throughout our experiments.

First, we designed 55~nm thick free-standing photonic crystal slabs (lattice constant 209 nm) with so-called L3 cavities~\cite{Akahane2003}  using FDTD simulations (\textit{Lumerical}). 
These cavities are formed by three missing holes in a triagonal lattice. 
They support several spectrally narrow modes of high quality factors~\cite{Chalcraft2007}. 
The electromagnetic field of the fundamental mode is concentrated mainly within GaP on about seven unit cells, corresponding to a material volume of 14~attoliter (see Fig.~1).
The emission profile of this fundamental cavity mode is polarized perpendicular to the cavity axis~\cite{Barth2008}, a feature that we will use later on to measure the cavity resonance and demonstrate optical switching.
The spectral position and Q-factor of the mode depends not only on the geometry (lattice constant, hole radius, slab thickness), but also on the refractive index of the PC material~\cite{Joannopoulos2007}.
\begin{figure}[htb]
\centering
 \includegraphics[width= 0.5\columnwidth]{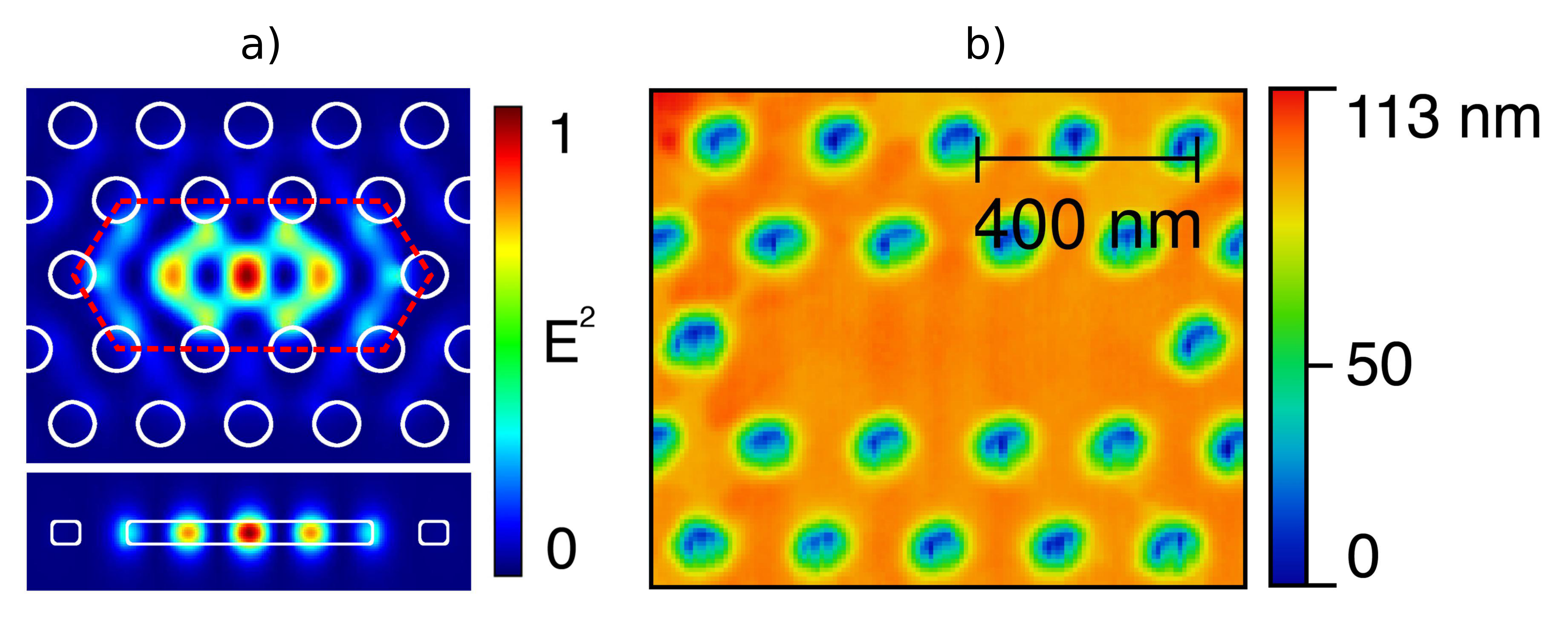}
  \caption{a) Simulated (FDTD) electric field profile in the slab center and cross-section of the fundamental mode of the cavity.
 The dashed red line indicates the seven unit cells in which the field is concentrated.
  b) Atomic force micrograph of the L3 cavity in GaP. The lattice constant is 209 nm. 
}
  \label{fig:1}
\end{figure}
%

For example, changing the refractive index of the  PC membrane by about 100~ppm
corresponds to a wavelength shift by $(54\pm1)$~pm, which can be detected by a spectrograph and is relevant for experiments with high-Q cavities~\cite{Englund2007}.
FDTD simulations unravel that in the studied regime the change of the resonance with the refractive index does not depend on the thickness of the photonic crystal slab (Fig.~2). 
Thus, a change in the resonance wavelength is a very precise measure for changes in the refractive index, even if the exact geometry is not known.
In contrast to other methods allowing refractive index measurements in bulk material~\cite{Bertolotti1990} or thin membranes~\cite{Cocorullo2002}, this allows measurements in ultra small volumina on the order of 10 attoliters. 
To the best of our knowledge this is the smallest volume in which refractive index measurements were ever performed.

We fabricated the structures by electron beam lithography (EBL) using a Vistec~5000+ electron beam lithography system with 100kV acceleration voltage.
Therefor a 59~nm heteroexpitaxial gallium phosphide (GaP) layer was deposited on a silicon  (100)  substrate.~\cite{Doescher2008, Doescher2010, Doescher2011}.
The lithography step was done in a 2.2M~PMMA resist layer spincoated on the GaP layer and exposed by EBL with a dose of 700~$\mu$C/cm$^{2}$. 
The designed structure was transferred to the GaP layer by dry-etching with boron trichloride (BCl3) and a  subsequent removal of the underlying Si layer by isotropic dry-etching with sulfur hexafluoride (SF6). 
After the fabrication process the hole sample was cleaned in an oxygen plasma using a Oxford Plasmalab 80.
Importantly due to a tiny lattice mismatch, the GaP layer is strained and may bend when it is underetched. 
To allow for expansion without bending, we designed spring-like membrane supports (see Fig.~\ref{fig9}(a)).

After processing, the quality and geometric precision of the structures were measured using scanning electron and atomic force microscopy (AFM). Fig~1(b) shows an AFM image of the fabricated structure. 
\begin{figure}[htb]
\centering
  \includegraphics[width= 0.5\columnwidth]{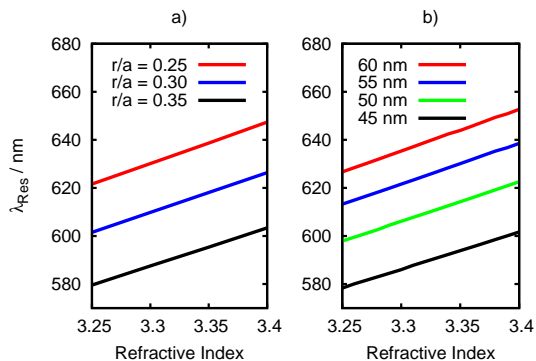}
  \caption{Simulated dependence of the first cavity resonance wavelength on the refractive index for various hole radii $r/a$ (a) at fixed $t = 55$~nm and slab thicknesses $t$ at fixed $r/a=0.3$ (b). The behavior is nearly linear and the slope does not depend on the thickness of the slab or on the hole diameter in the studied regime. 
}
  \label{fig:2}
\end{figure}
\section{Refractive index measurement}
\subsection{Experimental methods}
To control the cavity temperature within a wide range, the sample was mounted on the cold finger of a continuous flow He-cryostat.
By changing the He-flow and additional heating the temperature could be precisely regulated in the range between 5~K and 300~K. In parallel, the spectral position of the cavity resonance was measured. 
Usually, the intrinsic material fluorescence under incoherent excitation is analyzed with a spectrograph for this purpose~\cite{Barth2007}. 
Here, this method could not be applied, as the excitation of fluorescence requires relative strong lasers (in the order of $10~\mu W$), which would unavoidably heat the PC cavity. 
To avoid this heating, we analyzed the reflected light in the crossed polarization scheme~\cite{Galli2009}.
In this scheme only very low illumination power is needed ($P<100$~nW). 
By doubling the power and comparing the position of the cavity resonance we could guarantee that the radiation has negligibible influence on the cavity temperature ($<$~1~K).

\begin{figure}[htb]
\centering
  \includegraphics[width= 0.4\columnwidth]{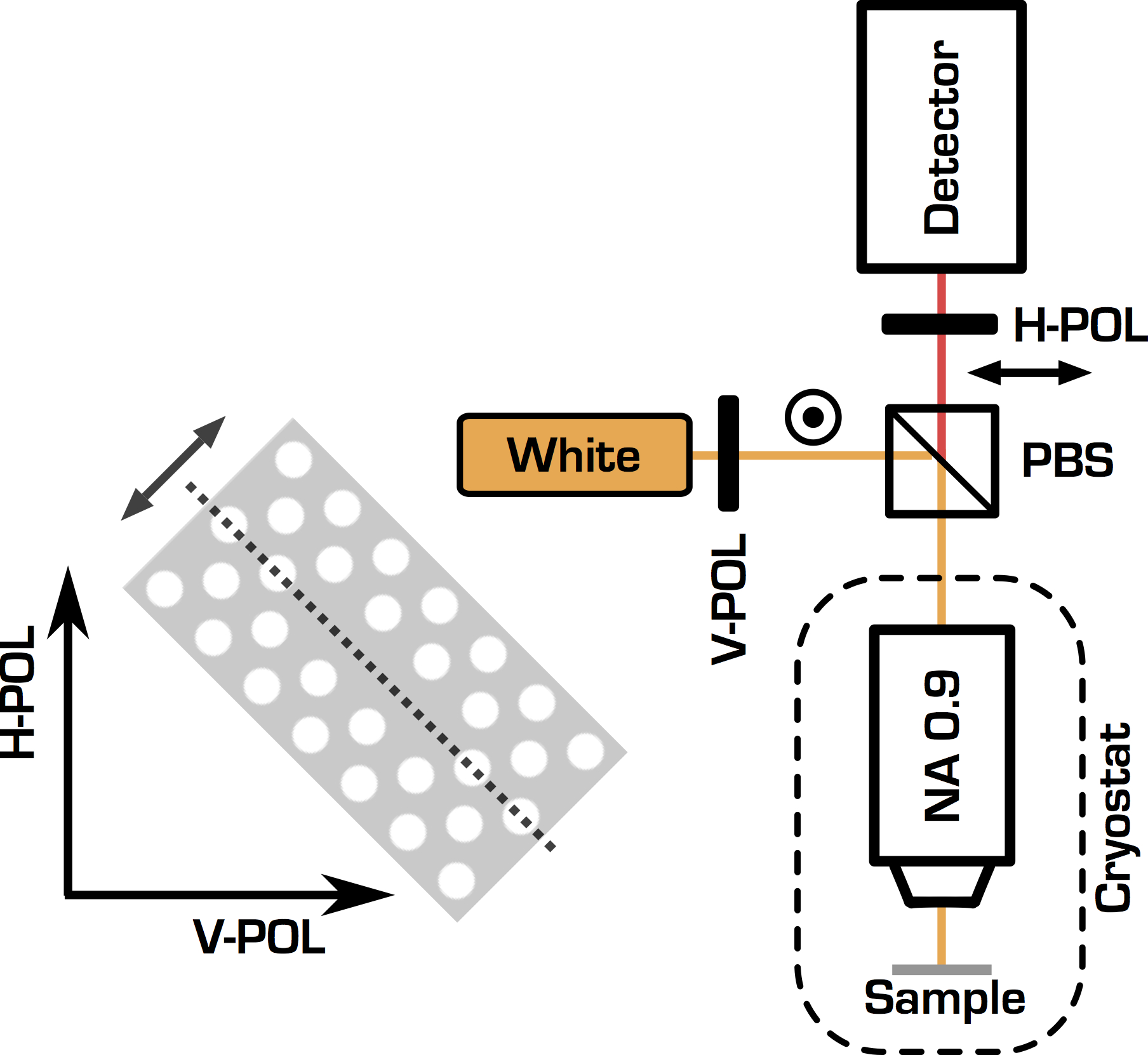}
  \caption{Sketch of the optical setup to measure the cavity resonance in the cross polarized detection scheme. The sample with the PC cavities is placed in the vacuum of a continuos flow He-cryostat to regulate the temperature. Through an objective lens with high numerical aperture the PC cavity is illuminated with vertical polarized white light ($\approx$100 nW) from a supercontinuum source. The horizontal polarized component of the reflected light is detected in a confocal scheme by a spectrograph. The cavity axis is rotated by 45 degree with respect to the polarization basis. Thus, the polarization of the reflected light is slightly rotated at the cavity resonance. See text for details.
}
  \label{fig:setup}
\end{figure}

In detail, the cavity is illuminated with a vertical polarized collimated white light beam from a super continuum source (\textit{NKT Photonics SuperK}) focused on the cavity by a microscope objective with a numerical aperture of 0.9 (\textit{Mitutoyo}), which is placed inside the insulation vacuum of the cryostat. Importantly, the axis of the cavity is rotated by 45 degree  with respect to the incident polarization. 
The reflected light is collected through the same objective lens, but only the horizontal polarized component of the reflected light is detected with a 500 mm spectrometer (\textit{Acton SpectraPro 500i}) in a confocal configuration.  A sketch of the setup is shown in Fig.~\ref{fig:setup}.
If the incident light is not in resonance with the cavity, the beam is simply reflected and can not be detected by the spectrometer. 
If the light is in resonance with the cavity, there are two effects  which slightly rotate the polarization and therefore can be detected on the spectrometer. 
First, the fraction of incident resonant light which is polarized along the cavity mode (perpendicular to the cavity main axis) couples into the cavity mode.
Then the reflected light has a horizontally polarized component which can pass the polarizing beam splitter and reaches the spectrometer.
Second, light that was stored in the cavity is re-emitted after the characteristic cavity decay time. 
This light also has a horizontal component. 
Both components then interfere at the detector and lead to the observation of a Fano-type resonance
\begin{figure}[htb]
\centering
  \includegraphics[width= 0.4\columnwidth]{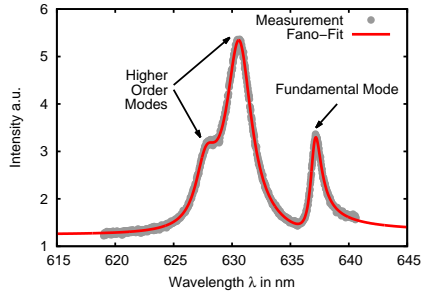}
  \caption{Example of the measured spectra and the corresponding fit with three Fano resonances at ($637.03\pm 0.01$)~nm, ($630.59\pm 0.01$)~nm and ($627.68\pm 0.01$)~nm of an L3 cavity with hole radius $r = 50$~nm. The spectrum was taken at a temperature of 300~K. 
}
  \label{fig:4}
\end{figure}
of the form: 
\begin{eqnarray}
F(\lambda)&=&A_{0}+F_{0}\frac{\left[q+2(\lambda-\lambda_{0})/\Gamma\right]^{2}}{1+\left[2(\lambda-\lambda_{0})/\Gamma\right]^{2}},
\end{eqnarray}
where $A_{0}$ is an offset, $F_{0}$ is the amplitude of the resonance, $\lambda_{0}$ is the resonance wavelength, $\Gamma = \lambda/Q$ the normalized resonance width and $q$ the ratio of the amplitudes of the two paths\cite{Fano1961}. 
Fig. 4 shows an example spectrum of a cavity with a hole radius of 50~nm and the fundamental resonance at about 637~nm, taken at room temperature.
The spectrum can be perfectly fitted with three Fano-resonances. The fundamental mode supports the highest Q-factor of \mbox{$Q=580\pm50$}. With the value from Ref.~\cite{Landolt} for the refractive index at $T=300$~K the slab thickness is estimated according to Fig.~2 to be 55~nm. 

\subsection{Temperature dependency of the refractive index of GaP and the cavity resonance}
In the following, the resonance wavelength of the fundamental mode of a cavity with hole radius $r=63$~nm fabricated on the same sample was measured at various temperatures between 5~K and near room temperature. To verify the reproducibility, the data were taken in two subsequent cooling and heating cycles. We calculated the refractive index $n$ from the change of the resonance wavelength according to the relation
\begin{eqnarray}
n(T) &=& n_{300K} + \left[\lambda(T) - \lambda_{300K}\right]\cdot a_{55},
\end{eqnarray}
with the refractive index at room temperature $n_{300K}=3.34537$ from Ref.~\cite{Landolt}, the measured resonance wavelength at room temperature $\lambda_{300K}=(608.20\pm0.01)nm$ and the slope $a_{55}=1/(163.36\pm0.01)$nm gained from the FDTD simulations for a slab thickness $t$ of 55~nm (Fig.~2). The resulting data and fitted curves are shown in Fig.~5. 
\begin{figure}[htb]
\centering
  \includegraphics[width= 0.5\columnwidth]{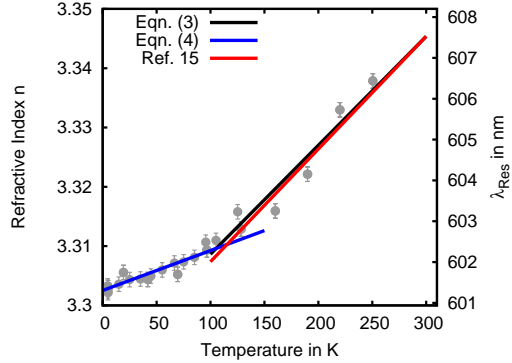}
  \caption{The measured cavity resonance wavelength $\lambda_{{Res}}$ (right axis) and the calculated refractive index of GaP (left axis)  between 5~K and 300~K. To verify the reproducibility, the data were taken in two repeated cooling and heating cycles.
The data above 100~K are in good agreement with Ref.\cite{Pikhtin1977}.
}
  \label{fig:5}
\end{figure}
Above 100~K the refractive index changes linearly according to 
\begin{eqnarray}
n(T) &=&3.290\pm0.001 - (180\pm20)\cdot10^{-6}\cdot T/\textrm{K}.
\end{eqnarray}
This is in very good agreement with values for the refractive index of GaP at temperatures above 100~K as reported in Ref.~\cite{Pikhtin1977}. Interestingly, the slope decreases at a temperature of $T\approx 100$~K and the refractive index follows
\begin{eqnarray}
n(T) &=&3.302\pm0.001 - (67\pm7)\cdot10^{-6}\cdot T/\textrm{K}.
\end{eqnarray}
We tried to fit the entire dataset shown in Fig.~5 with an exponential function, but the two slopes model represented by Eqns.~(3-4) is clearly more accurate. Although to the best of our knowledge no low temperature data for the optical refractive index is available in the literature, the kink at about 100~K is qualitatively in agreement with measurements of the low frequency dielectric constant~\cite{Samara1983}.

\subsection{Influence of the temperature on the quality factor}
\begin{figure}[tb]
\centering
  \includegraphics[width= 0.5\columnwidth]{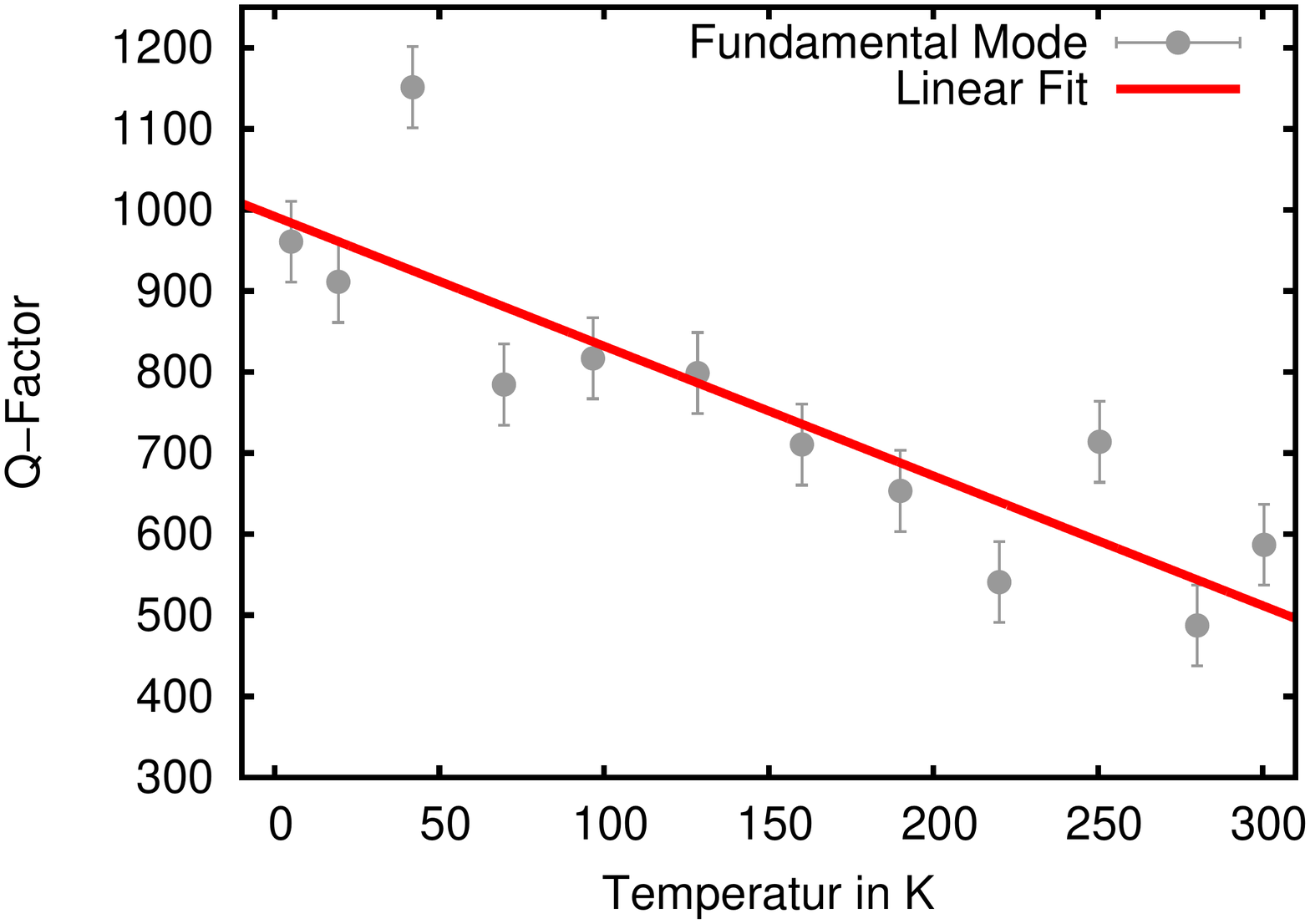}
  \caption{The measured $Q$-factor of the fundamental cavity resonance increases with decreasing temperatures.
   The highest measured value of $Q=1150$ at $T=42$~K is about two times bigger than the value at room temperature. 
}
  \label{fig:6}
\end{figure}
According to FDTD simulations the $Q$-factor of  $Q \approx 2500$ is almost unaffected by a small change of the refractive index.
We find experimentally that the decrease of the refractive index is followed by an increasing Q factor (Fig.~6). 
Starting with a value of $560\pm50$ at room-temperature, the $Q$-factor of the fundamental mode increases linearly with decreasing temperature. The highest measured $Q$-factor of $1150\pm50$ is more than two times higher than the initial value at room-temperature. 
We attribute this behavior to reduced material absorption~\cite{Dean1966, Subashiev1966} at low temperature due to freeze out of phonons. 

This dependency highlights the importance of considering not only the desired operation wavelength, mode volume and $Q$-factor, but also the operation temperature of photonic crystal cavities.

\section{Thermo-optical switching}
The shift of the PC cavity resonance as a function of temperature can also be used to implement integrated thermo-optical switching. Such switches have been demonstrated in the infrared. There, a modification of the index of refraction via optical free carrier injection [30, 31, 32] is preferable due to the faster switching speed needed in telecom technology.
In the following we exploit switching in the visible spectral range. Since the size of integrated optical devices scales with the third power of the wavelength cavity-based switching can be achieved with much smaller cavity volumes. As we will show this in principle leads to fast switching speeds even via thermo-optic effects.
\begin{figure}[t]
  \centering
  (a)\hspace{6cm}(b)\\
 \includegraphics[height=4.5cm]{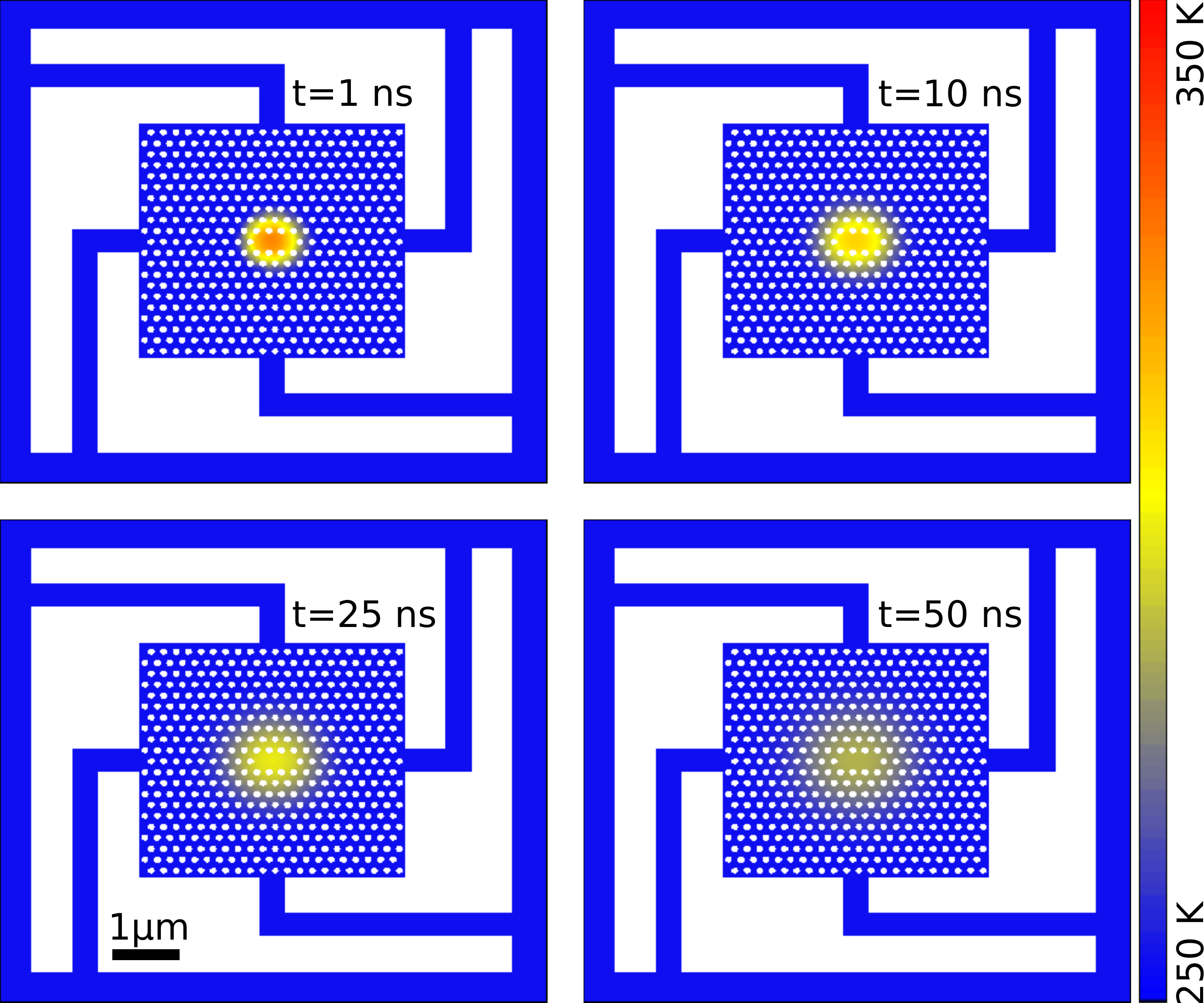}
  \includegraphics[height=4.5cm]{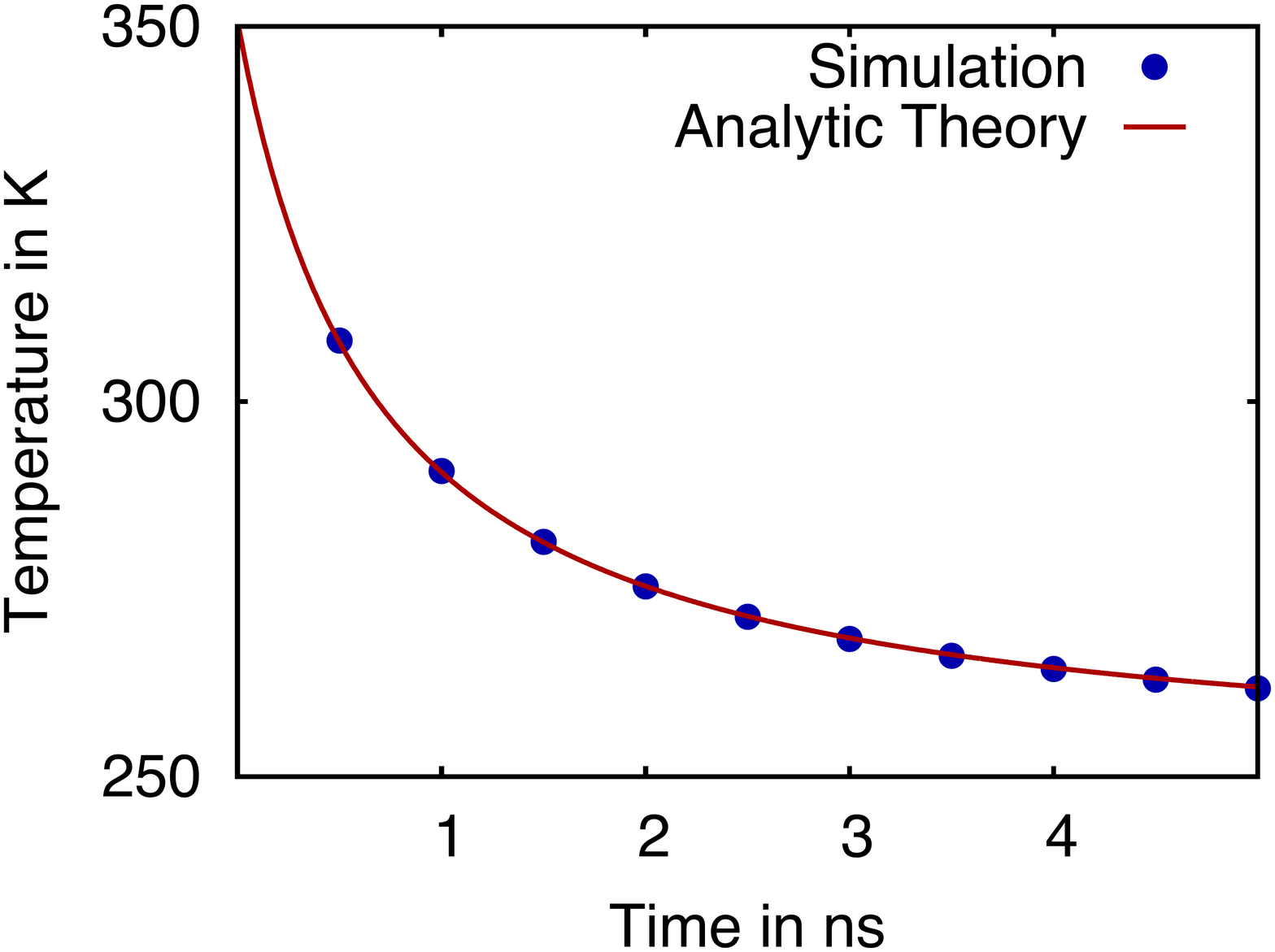}
  \caption{(a) Snapshots of the simulated temperature distribution 1 ns, 10 ns, 25 ns and 50 ns after injection of a Gaussian shaped heat pulse (b) Cavity temperature after a delta-shaped pulse. 
  The figure shows the result of the simulations (dots) and the analytical solution (solid line) with effective thermal diffusivity.}
  \label{fig9}
\end{figure}
\subsection{Theoretical predictions}
To estimate the energy required to switch the device, we assume a specific heat of 410~J/(kg K) \cite{Merill1977,Tmar1984} at operation temperature of 250~K.
Assuming that the cavity volume is heated by a short laser pulse and using our experimental findings from Section 4 we find that the cavity resonance shift directly after the laser  pulse is 1.3~pm per fJ of deposited energy.
For cavities with moderate Q-factors of $10\,000$ the temperature induced resonance shift is on the order of the width of the resonance, when depositing 48~fJ.
Thus thermo-optical switching with less than 50~fJ switching energy is feasible. To further reduce the energy required for switching, the operation temperature might be further reduced.

In the device thermal radiation can be neglected and the dynamic behavior of the temperature $T$  is governed by  the heat equation
\begin{eqnarray}
\frac{\partial}{\partial t}T(\mathbf x,t)&=&\alpha(\mathbf x) \triangle T(\mathbf x,t)
\end{eqnarray} 
where the thermal diffusivity $\alpha(\mathbf x)$ at position $\mathbf x$ equals the diffusivity in GaP $\alpha_{GaP} = k/ \rho c_p = 66\times 10^{-3}\,\mu$m$^2/$ns \cite{Weil1968} in the membrane, while it is zero in the holes.
To investigate the switching dynamics, we performed 2-dimensional finite-difference simulations of the heat transport within the photonic crystal membrane and the mechanical support.
In detail, the problem was tackled using the Crank-Nicolson method \cite{Crank1947} and the remaining sparse system was solved by the conjugate gradient method \cite{Hestenes1952}.
Here, Drichlet boundaries were used to implement the heat reservoir of the bulk material, whereas Neumann boundaries were used to simulate the PC holes which inhibit the heat conduction process. 
Fig.~7(a) shows the resulting temperature distribution in the PC slab at different times after the injection of a 3.7~pJ heat pulse.
It is clearly visible, that in the case of a single pulse the comparable small thermal conductivity of the membrane holders has no effect.
To estimate the switching speed, the temperature evolution at the cavity position was evaluated (Fig.~7(b)).
This temperature evolution can perfectly be fitted with the analytical solution of the heat equation for a two dimensional slab
\begin{eqnarray}
T(t) &=&\frac{1}{4\pi\alpha_{eff} (t-t_0)},
\label{eq:2ddiff}
\end{eqnarray} 
where the effective diffusion coefficient $\alpha_{eff}=(42.61 \pm 0.04)\times 10^{-3}\,\mu$m$^2/$ns.
Remarkably this can not be explained by an average diffusivity deduced from the PC filling factor $\beta=0.8$, which is $\beta \alpha_{GaP}=53\times 10^{-3}\,\mu$m$^2/$ns.
Furthermore, according to Eq.~(\ref{eq:2ddiff}) the cavity temperature difference is decreased to $1/e$ of the initial value after the time $\tau = 1.2$~ns.
Thus, the device promises switching speeds an the order of 1~ns.
In contrast for longer pulses, where the dynamics is limited by the heat transport through the support springs we find $\tau \approx 680$~ns.
To increase the switching speed  the structure might be partially covered with a thin gold layer to increase the thermal conductivity and thereby allow for faster switching.
Furthermore the operation temperature might be further decreased, resulting in an improved ratio between specific heat and diffusivity.

\subsection{Experimental implementation}
To experimentally verify our simulations, we pumped the PC cavity material with a switchable blue 405~nm laser (PicoQuant LDH-P-C-405B). 
The light is efficiently absorpt and thus heating the cavity. 
Simultaneously while switching the heating laser, we probed the cavity resonance as described in section 3.1. 
We used the spectrometer as a monochromator and detected the light at the cavity wavelength with a fast avalanche photo diode.
This allowed for high time resolution of up to 1~ns.

\begin{figure}[htb]
\centering
  \includegraphics[width= 0.4\columnwidth]{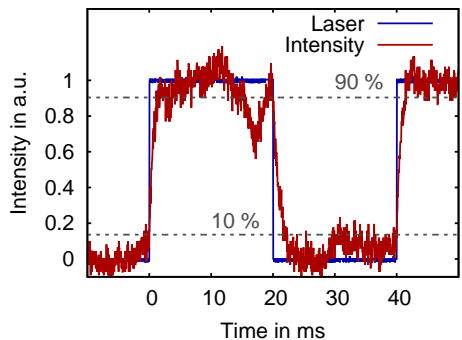}
  \caption{Real time measured reflected intensity at the resonance wavelength when applying the heating laser for 20~ms intervals. 
  The achieved contrast between on-state and off-state is above 13~dB, while the 10\% to 90\% switch-on time is 1.2~ms. Switch-off is 2.2~ms. Negative intensities are due to correction for the detector background.
}
  \label{fig:8}
\end{figure}

The observable contrast when switching the heating laser on and off exceeded 13~dB.
Nevertheless, we could only observe switching times on the order of about 1~ms (Fig. \ref{fig:8}) for laser powers of about 73~$\mu$W. 
This is over three orders of magnitude slower than predicted.
Furthermore, the required heating power is several orders of magnitude above the predicted value.
The large deviation originates in the parasitic heating of the entire substrates:
A major fraction of the heating laser is absorbed in the silicon substrate, rather than in the membrane.
Thus the whole sample is heated up, which requires more energy and is much slower.
Using transparent substrates or a different heating mechanism, these problems might be solved in future experiments.
Nevertheless, already the current configuration is suitable to perform active control of the membrane temperature \cite{Englund2007}.

\section{Conclusion}
In conclusion, we demonstrated how the narrow resonance of a GaP photonic crystal cavity shifts due to cooling.
 From this data we precisely calculated the refractive index of GaP in the ultrasmall cavity volume over the whole temperature range from near room temperature to $T=5$~K.
 This knowledge is crucial for the design and testing of PC structures and might also be used for novel PC applications.
 For example, the presented method may be used to determine the exact local cavity temperature. This is of great importance if quantum emitters are coupled to the cavity, as the emission properties of such emitters strongly change with the temperature. 
In most previous experiments the temperature of the cold finger of the cryostat was given as a temperature reference. 
However, the cavity structure may be weakly coupled thermally to the cold finger and excitation light may locally heat the cavity. 
This is particularly an issue when studying under-etched photonic crystal membrane cavities and diamond defect centers as emitters, which require a relatively large excitation power~\cite{Wolters2010,Wolters2012}.
Finally, the index of refraction can be measured via our method for any material where fabrication of a resonant photonic structure is possible.

In the second part of the paper we investigated GaP photonic crystal cavities for thermo-optical switching applications.
By numerical simulations, we found switching speeds of up to 1~ns, and outlined strategies to further improve this value.
Furthermore, we realized a thermo-optical switch with 13 dB contrast between on- and off-state. \\

\section*{Acknowledgements}
This work was supported by the DFG (BE2224/9).
J.~Wolters acknowledges funding by the state of Berlin (Elsa-Neumann). We thank H. D\"oscher and T. Hannappel for providing GaP on Si wafers.
We thank PicoQuant for collaboration and support.

\end{document}